\documentclass[12pt]{article}

\usepackage{psfig}

\newcommand{\be}{\begin{equation}}
\newcommand{\ee}{\end{equation}}
\newcommand{\ba}{\begin{eqnarray}}
\newcommand{\ea}{\end{eqnarray}}

\newcommand{\DK}[1]{\mbox{\boldmath$#1$}}
\newcommand{\BE}{\begin{equation}\label}
\newcommand{\BEQ}{\begin{eqnarray}\label}
\newcommand{\EE}{\end{equation}}
\newcommand{\EEQ}{\end{eqnarray}}

\begin{document}

\title{Quasiclassical Theory and Simulations of Strongly Coupled Plasmas}

\author{W.~Ebeling\thanks{email:werner@summa.physik.hu-berlin.de}, J.~Ortner}

\date{}
\maketitle
\begin{center}
{Institute of Physics, Humboldt University Berlin,\\
Invalidenstr. 110, 10115 Berlin, Germany}
%\\(to appear in Phys.Scr.)
\end{center}

\begin{abstract}  

A survey on the dynamical and thermodynamical properties of plasmas with
strong Coulomb interactions in the quasi-classical density-temperature
region is given. First the basic theoretical concepts describing nonideality are discussed. 
The chemical picture is introduced.
It is shown that the nonideal plasma subsystem of the free charges has a rather large
quasi-classical regime, where the quantum effects yield only
corrections to the merely classical dynamics.
The plasma of free charges may be described by 
effective potentials which incorporate quantum effects in an approximative way. 
The simplest effective potentials are
only space-dependent, more advanced methods include momentum-dependent 
interactions. On the basis of these
potentials analytical results are derived and simulation methods are developed. 
It is shown that effective potentials are appropriate for the description
of thermodynamical as well as collective properties.

\hspace*{0.5cm}

\end{abstract}

\section{\protect\large{Introduction}}

Strongly coupled plasmas play an important role 
in nature, laboratory experiments, and in 
technology \cite{gruenes-buch,braunes-buch,ichimaru92,ksruegen}. In these 
plasmas the mean potential energy is of the same order of magnitude as the mean  
kinetic energy. Then we speak also about nonideal plasmas.
We will study in this work the dynamics, thermodynamic properties and several collective effects of strongly coupled non-degenerate
one-component plasmas (OCP) and symmetrical two-component plasmas (TCP). The investigation 
is restricted to the subsystem of the
free charges, which is defined by means of the chemical picture \cite{braunes-buch}.
In this model the atoms, ions, and molecules are treated as separate
species. Therefore the constituents of the plasma are free electrons, free 
nuclei, ions, atoms, and molecules. All species are treated on equal footing
(principle of particle democracy). The advantage of the chemical picture is
that it is in many cases more appropriate for the description of real 
plasmas \cite{braunes-buch}. In the chemical picture the free charges constitute a 
well defined subsystem of the plasma. 

The main purpose of this work is the investigation of a quasi-classical dynamics 
of free charges based on effective potentials
which may be used to describe thermodynamical and collective
properties of the free charges in this subsystem. We restrict ourselves to
the classical and near-classical region. Degeneracy is taken into account only in an 
approximative way. Therefore the new results refer in particular to a region
where the plasma is still nondegenerated
but nevertheless strongly coupled \cite{ebeling}.
In this work we will restrict ourselves to one-component plasmas and to two-component 
plasmas which are anti-symmetrical with respect to 
the charges ($e_- = - e_+$) and symmetrical with respect to the densities
($n_+ = n_i = n_- = n_e$). In particular we consider
the model case of electron - positron plasmas $m_+ = m_e$ and H-plasmas with 
the mass-relation $m_+ = 1840~m_e$. We include the - so far unrealistic - case 
of mass - symmetrical plasmas since in this case the thermodynamic functions 
and other analytic functions describing the plasma
are of particular simplicity. This is due to cancellation effects caused by the symmetry
of the masses and anti-symmetry of the charges \cite {LeEb96}.
The effective potentials may be used to calculate correlation functions, 
thermodynamic properties and structure factors of the free charges in
semi-classical non-degenerate quantum plasmas. 
The effective potentials are first obtained from the Slater sum method. 
Then momentum-dependent potentials are introduced and discussed.

{\em Characteristic parameters: }
The average distance of the electrons is the Wigner-Seitz radius 
$ d = [3/4\pi n_e]^{1/3} $, 
the Bohr radius is defined as
$ a_B = \hbar /me^2 $.
Other characteristic lengths are the 
Landau length $l = e^2 / kT$, the 
De Broglie wave-length $\Lambda_i = h/[2\pi m_i kT]^{1/2}$ of particles
of species $i$ and thermal wave length of relative motion
$\lambda_{ij} = \hbar/(2 m_{ij} kT)$. 
Furthermore we define 
the following dimensionless parameters:\\
(i) the coupling strength: $\Gamma = l/d $. \\
(ii) the degeneration parameter: $n_i \Lambda_i^3$ or $\theta=2mkT/[\hbar^2 \, (3 \pi^2 n_e)^{2/3}]$.\\
(iii) the interaction parameter $\xi_{ij} = - (e_i e_j)/(kT \lambda_{ij}$).\\

Let us first discuss the OCP:
In the limit when the average distance is large 
in comparison with the Bohr radius, i.e. if $d \geq a_B$  the
electron gas behaves  classical.
The classical case was treated analytically by 
Ab{\'e} and others \cite{CoMu69}, Monte Carlo calculations in a wide range of
 $\Gamma$-values were carried out e.g. by Brush, Sahlin and Teller,
DeWitt, Ichimaru and other workers 
(see  \cite{march-tosi,gruenes-buch,ichimaru92,ksruegen}).
Quantum corrections to the classical case which are relevant
at moderate values of $r_s$ were investigated by many authors
\cite{gruenes-buch,hoffmann_ebeling,deutsch}.
All the analytical calculations mentioned so far cover only limiting
cases as e.g. $r_s < 1$ or $\Gamma < 1$. This is the reason
why simulations are of large interest. \\
There exists extensive Monte Carlo (MC) calculations for the classical
region \cite{dewitt76}. Classical molecular dynamics (MD) calculations were presented by Hansen et al. \cite{hansen}. 
The particular interest in MD calculations is connected
with the fact, that they give us also access to non-equilibrium
properties \cite{hansen}.

For the TCP the available material is less exhaustive. We mention analytical calculations 
in the region of small $\Gamma$ 
\cite{Eb67,ebeling,ekk70,alastuey2}
and Pad{\'e}-approximations \cite{braunes-buch,ebelingcontr}
connecting the analytically accessible regions.
Further we mention several investigations
devoted to the simulation of two-component plasmas
\cite{norman,hansen2,penman95,PC94,KTR94,EbMiSc97}.
In particular our interest is devoted here to quasi-classical methods. Quasi-classical 
simulations of two-component plasmas were pioneered by Norman and Hansen 
\cite{norman,hansen2}.
These methods attracted a great deal of interest because of their relative
simplicity. A special direction in developing simulation methods uses
momentum-dependent potentials. The main idea in this approach is to model
quantum effects by certain constraints in the phase space constructing an appropriate Hamiltonian \cite{He75,Dorso2,FBS95}.
In particular the Pauli exclusion principle is simulated by a momentum-dependent
two-body interaction.

\section{Effective Interaction Potentials}     

\subsection{Space - Dependent Effective Interactions}

As pointed out the idea of quasi-classical methods is to incorporate
quantum-mechanical effects (in particular the
Heisenberg and the Pauli principles) by appropriate potentials.
Such a quasi-classical approach has of course several limits which
are basically connected with the trajectory concept. We mention
for example the principal difficulty to describe microspic quantum
effects as tunnelling, and macroscopic quantum effects as superfluidity
and superconductance. Our aim is only the calculation of standard
macroscopic properties which have a well defined classical limit.
Since bound states cannot be described classically our methods are 
restricted to the subsystem of free charges. However, this is not a very serious
restriction since most of the plasma properties are determined by the subsystem
of the free charges.\\
The easiest way to arrive at effective potentials describing quantum effects is the use of
the so-called Slater sums which are defined by the N - particle wave functions,

\be
S( \DK{r}_1,\ldots,\DK{r}_N ) = \mbox{const} \sum \exp \left(-\beta\;E_n \right)\;
\left|\Psi_n\left( \DK{r}_1,\ldots,\DK{r}_N \right)\right|^2 \quad .
\ee

The integrals over the distributions $S( \DK{r}_1,\ldots,\DK{r}_N )$ 
yield the correct quantum statistical partition function. The Slater 
sums for Coulombic systems
were studied in detail by several authors \cite{gruenes-buch,ebeling}. 
With the knowledge of the Slater sums one gets exact space distributions in equilibrium by the choice

\be
U^{(N)}(\DK{r}_1,\ldots,\DK{r}_N ) = - k T \ln S( \DK{r}_1,\ldots,\DK{r}_N ) \quad .
\label{slater}
\ee

These potentials are often called quantum statistical effective potentials
and they are used to calculate the partition function \cite{gruenes-buch,ebeling,norman}. Thus one obtains the correct thermodynamic functions. However, a severe disadvantage of this choice of the effective
potentials eq.(\ref{slater}) is, that the momentum distributions are correctly described  only in the
Boltzmann limit. Therefore this approach is principally restricted to the
non-degenerate region.

The Slater sum is an analogue of the classical Boltzmann factor and one
defines therefore an effective potential by

\BE{slater-sum}
 S_{ab}^{(2)} (r) = \exp \left( -\beta u_{ab}(r) \right)
   = const. \sum_\alpha \exp\left( -\beta E_\alpha \right) 
   \mid \Psi_\alpha \mid^2 \;\;.
\EE

Here $\Psi_\alpha$ and $E_\alpha$ denote the wave functions and energy
levels of the pair $ab$.
 
A quantum mechanical calculation including the first orders in the perturbation theory was first given by Kelbg; a similar more simple effective potential was 
derived by Deutsch and was used in the simulations by Hansen and McDonald 
\cite{hansen2}. \\
The effective potentials derived from perturbation theory do not include bound 
state effects. 
In order to treat the region, where bound states are of importance, a quite different approach is
necessary \cite{ebeling}. The first step is a transition to the chemical picture i.e. 
bound and free states have to be separated. The second step is the derivation of an effective 
potential for the interaction of the free charges in the plasma.
In order to proceed we split the Slater sum into a bound part
and a free part. This splitting is not unique. We use here the so - called
Brillouin - Planck - Larkin (BPL) convention.
This way of division free - bound corresponds to a smooth cut - off of the partition 
function which is based
on omitting of the divergent elements in the sum. This leads to \cite{gruenes-buch}
  
\BE{bound-part}
  S_{ab}^b(r) 
   = const. {\sum_\alpha}'
    \left( \exp\left( - \beta E_\alpha \right) -1 + \beta E_\alpha \right)
    \mid \Psi_\alpha \mid^2  \; ,
\EE

where the sum extends only over all the discrete states. It is well known, that the BPL -
convention leads to the simplest expressions for the thermodynamic functions. 
The Slater sum of the free charges is defined by
\cite{gruenes-buch,Eb67,ebeling}:

\BE{free-part}
  S_{ab}^* (r) =  S_{ab}^{(2)} -   S_{ab}^b
   = \exp\left( - \beta u_{ab}^*(r) \right) \quad .
\EE

In this way we arrive at an effective interaction potential of the free
particles $u_{ab}^*(r)$, which is finite for $\hbar \neq 0 $ and has a weak
(integrable) singularity in the limit $\hbar \rightarrow 0 $.

For the electron - electron and for the ion - ion  interaction the classical
limit gives simply the Boltzmann factors, i.e.
$u^*_{ee} = u^*_{ii} = e^2 / r $.
For the electron - ion pairs one has to perform the classical limit
in eq.(\ref{free-part}).
Explicitly one finds for the classical limit within the 
BPL - convention \cite{ebeling,ksruegen}:   

\BEQ{s-ei}
 S^*_{ei}(r) &=& \exp \left(\frac{Z e^2}{kT r} \right) \cdot 
      \left[ 1- \Phi \left( \sqrt{\frac{ Z e^2}{kT r} } \right) \right]
    + \frac{2}{\sqrt{\pi}} \sqrt{\frac{Z e^2}{kT r} }
    \nonumber \\ 
 &&   + \frac{4}{3\sqrt{\pi}} \left(\frac{Z e^2}{kT r} \right)^{3/2}
    + \frac{8}{15\sqrt{\pi}} \left(\frac{ Z e^2}{kT r} \right)^{5/2} \; \;,
\EEQ

where $\Phi(x)$ denotes the error function. The effective potential

\be
u^*_{ab}(r) = - kT \cdot \log ( S^*_{ab}(r))
\ee

may be used for simulations of the free charges in the purely classical region.\\

\subsection{Momentum - Dependent Effective Interactions}

 As mentioned already above, a principal disadvantage of purely space-dependent potentials is
the incorrect representation of the momentum - distributions of the plasma.      
In order to achieve a correct representation of
the Fermi distribution for the momenta,
momentum-dependent potentials have to be included \cite{Dorso2,KTR94,EbSc97}.
In what follows we will assume a quasi-classical Hamiltonian of the following structure:

\be
H=\sum_{i=1}^N \frac{p_i^2}{2m} +
\sum_{i<j} V_P \left( \frac{r_{ij}}{r_{ij0}}, \frac{p_{ij}}{p_{ij0}} \right)
+ \sum_{i<j} \frac{e^2}{r} \cdot F \left( \frac{r_{ij}}{r_{ij0}},\frac{p_{ij}}{p_{ij0}} \right) \quad .
\label{qcHam}
\ee

Here $r_{ij}$ is the usual distance in the coordinate space and $p_{ij}$ the distance
in the momentum space. Further we define the characteristic parameters
\be
r_{ij0}^2 = r_{i0}^2 + r_{j0}^2; \qquad  p_{ij0}^2 = p_{i0}^2 + p_{j0}^2; 
\qquad p_{i0} = \hbar / r_{i0} \quad ,
\ee

where $r_{i0}$ is a characteristic length (i.e., the radius of the wavepacket) of the 
particle $i$.
We have in the Hamiltonian two kinds of particle interaction:
the so-called Pauli-potential $V_P$ acting only between identical particles 
and a Coulomb interaction modified
by a certain function $F(x,y)$.
In order to derive effective expressions of this type 
the Hamilton operator $\hat {H}$ is averaged with respect to test wave functions \cite{KTR94}, 

\be
H(q,p;\hbar) = \int d{\bf x} \psi^*_o (x) {\hat {H}} \psi_o (x) \quad .
\label{Hbypsi}
\ee
This definition of an effective Hamiltonian stems from the so-called wave-packet dynamics. In the last time it has found several applications to plasmas
\cite{KTR94,ebeling_militzer,EbFoPo97,OrScEb97}.
If one chooses for the test wave functions symmetrized and anti-symmetrized combinations of minimum uncertainty wave packets
for the particles of species $k$,

\be
\psi_{k0} (x) = \mbox{const} \exp \left( - \frac{(x - q)^2}{ 2 r_{k0}^2} + 
\frac{i px }{ \hbar}\right) \quad ,
\label{wavepacket}
\ee

one does not get any Pauli potential for two electrons with antiparallel spins, for two electrons with parallel spins the following Pauli potential is obtained,

\be
V_{P_{ij}} (x, y) = \delta_{ij} \frac{\hbar^2}{m r_{ij0}^2} \; 
\exp \left( - \Delta ^2\right) \; \frac {\Delta^2 }{1 - \exp( - \Delta^2)} \quad .
\label{wp_av}
\ee

This is a two-body interaction depending on
the phase-space distance $\Delta^2 = x^2 + y^2$.
In our simulations we averaged over the two spin configurations and 
used a simplified Pauli-Potential which is
purely Gaussian \cite{Dorso2},

\be
V_P (p, r) = V_0 \cdot \exp(-\Delta^2) \quad , \qquad V_{ij0}=\frac{\hbar^2}{2 m\,r_{ij0}^2} \qquad \mbox{.}
\label{dorso_pot}
\ee

The averaging over wave functions leads also to a modified Coulomb interactions, described by the function $F$ (eq. \ref{qcHam}). We obtain

\be
F(x,y) = \mbox{erf}(x) \quad .
\ee

This potential was obtained already by Klakow et al. \cite{KTR94}, it represents the electrostatic energy 
between two charges which are Gauss-distributed. The free
potential parameter $r_0$ was fitted in such a way, that the properties of an
electron gas without interactions as the binary correlation function and the Fermi momentum distribution are well reproduced. We assume that the momentum uncertainty is given by that of the free Fermi gas.
%\begin{equation}
%(\delta p)^2 = 2 m \epsilon^{(i)}_{kin} \quad ,
%\end{equation}
%
%where $\epsilon^{(i)}_{kin}$ is the mean kinetic energy
%of the free Fermi gas of particles of species $i$.
In this way we get

\be
p_{i0}^2 = \frac{4}{3}\,m\, \epsilon^{(i)}_{kin} 
\ee

and $r_{i0} = \hbar / p_{i0}$, where $\epsilon^{(i)}_{kin}$ is the mean kinetic energy of the free Fermi gas of particles of species $i$.  These relations yield at high temperatures
$p_{i0} ^2 =  m_i kT; r_{ii0} = 2 r_{i0}^2  =  \hbar ^2 / 2 m_i kT = 
\lambda_{ii} ^2 / 2; V_0 =  kT / 2$. We mention that there exist other estimates for the free parameters of the Pauli-potential \cite{Dorso2}.

\section {Quantumstatistical Theory and Simulations} 

\subsection{Thermodynamic Properties}

Quantum corrections to the classical electron gas were derived in earlier
work \cite{hoffmann_ebeling,EbSc97}. Generalizing the methods developed earlier \cite{EbSc97} to the case of the TCP we assume that the interaction part of the
free energy density of the plasma
can be split into a classical and a quantum-mechanical part

\be
f_{int} = f_{cl}+ f_{qu} \quad ,
\ee

where $f_{cl}$ is the known free energy density of the
classical plasma of free charges and $f_{qu}$ is the difference between
the full and the classical free energy density.
Explicit calculations for the classical free energy of
the electron gas were given first by Ab{\'e} for the low density case and extended by 
Cohen and Murphy \cite{CoMu69}. For a symmetrical classical TCP it was shown
that (with the BPL  convention) the first correction beyond the Debye term in the density expansion of the free energy vanishes. \cite{Eb67}.\\ 
Consider now the quantum-mechanical corrections to the
classical free energy density. For the plasma of free charges 
this expression is convergent. In the low density limit we get

\be
f_{qu} = - kT \sum n_i n_j \delta B_{ij}(T) \quad .
\ee

Here $\delta B_{ij}(T)$ is the difference between the quantum and the
classical second virial coefficients.
This function can be calculated exactly by using methods of quantum scattering
theory \cite{ekk70}.
First the two particle trace is transformed to a
contour integral over the resolvent 
%which gives
%\be
%B_{ij} (T) = \frac{\pi^{3/2} \lambda^3}{2i} \int_C \exp (-\beta z) F_{ij}(z) dz
%\ee
%\be
%F_{ij}(z) = Tr \{ [ \hat{H_{ij}} - z ] - [ \hat{H_{ij}}^{0}- z ] \}
%\ee
%Here $\hat{H_{ij}}$ is the two-particle Hamiltonian and $\hat{H_{ij}}^0$ its kinetic part.Now the trace can be carried out by using the known Jost functions 
of the Coulomb scattering problem. For a mass-symmetrical TCP all terms except to one
cancel and we get (neglecting exponentially small degeneracy effects) the rather simple
result

\be
f_{qu} = - \pi^{3/2} n_i n_e \lambda_{ie}^3 \xi_{ie}^2 + O(n^{3/2}) 
\ee
If the masses are not symmetrical, infinite series in the $\xi$ - parameter appear, the
so - called quantum virial functions $Q$ and $E$ \cite{gruenes-buch,ebeling}.
The theory may be extended to higher concentration by using the method of
Pad{\'e} - approximations \cite{LeEb96}.

We carried out extensive simulations for the electron gas \cite{EbSc97} and 
also a few simulations for the symmetrical TCP by using
the effective Hamiltonian described above. 
Since our modified Coulomb interaction differs from the bare Coulomb
interaction only for
short distances the Ewald sum technique for handling the long range part
could be used.
To study equilibrium properties, as e.g. the average energy we carried out several Monte Carlo runs of $2\,\cdot \, 10^6$ steps for ensembles of 64-512 particles. The results were extrapolated to infinite particle
numbers. The regions of degenerate and nondegenerate plasmas were investigated. The results of our simulations were discussed in detail in earlier
work \cite{EbSc97,EbMiSc97}. As the main result we may quote, that at all densities the deviations from the classical calculations are very small
\cite{EbMiSc97}. 
Further we may state that the overall agreement between the analytical formulae and
the simulations is rather good. The deviations are all within the error bars of the
simulations. Only at low temperature a systematic deviation is observed, here the semi - classical
model does not work.
At conditions of weak or moderate
degeneracy our model yields quite reasonable results.

\subsection{Collective Excitations}

In order to study collective nonequilibrium effects we simulated a system of 250 electrons with periodic boundary conditions by Molecular Dynamics (MD) calculations. 
Usually MD simulations consist of two parts: First the desired temperature is adjusted using some
thermostat, while in a second phase the energy is kept constant and measurements are
performed. Since most known thermostats do not work in the case of momentum dependent
interactions we replaced the equilibration phase by a MC simulation
using the Metropolis Algorithm. This Algorithm is
independent from any particular form of the interaction.
To perform the MD simulations we used a 4th order Runge-Kutta
integrator with stepsize control. The runs were of length of order 1000 times
the inverse plasma frequency.

First the individual motion of the electrons was studied by calculating the 
velocity autocorrelation function $<v(t+\tau) v(t) >_t$ \cite{EbSc97,OrScEb97}.
It was shown that for moderate coupling ($\Gamma=1$) the velocity autocorrelation falls monotonically to zero , whereas for the case of strong coupling 
($\Gamma=100$) the velocity acf shows oscillations with a frequency close to 
the plasma frequency. 
%That means that in the strong coupling case the motion of a single electron is coupled to the collective density fluctuations.
 
To describe the collective behavior of the system we have investigated the dynamic structure factor of the electron system,
\begin{equation} \label{dynstruc}
S(\vec{k} ,\omega) \, = \, \frac{1}{2 \pi N} \int_{- \infty}^{\infty} \, 
e^{i \omega t} \, <\rho(\vec{k} ,t) \, \rho(- \vec{k} , 0)> \, dt \qquad ,
\end{equation}
where $\rho(\vec{k} ,t)=\sum_i \exp(-i\vec{k} \vec{r_i})$ is the Fourier component of the microscopic electron density.
We obtained the dynamic structure factor from the MD simulations by approximating the Heisenberg operator $ \vec{r_i} (t) $ by the position of the i-th particle in the simulations. However, the thus obtained quantity (we denote it by $ R( \vec{k} , \omega) $) is symmetric with respect to the frequency. It corresponds therefore to a classical fluctuation-dissipation theorem,
\begin{equation} \label{FDTcl}
R (\vec{k} , \omega) = (n \pi \phi (k) \beta \omega)^{-1} \, 
Im \, \varepsilon^{-1} (\vec{k}, \omega) \qquad .
\end{equation}
where  $ \varepsilon (\vec{k}, \omega)$ is the dielectric function of the electron OCP.
It can be seen from Eq. (\ref{FDTcl}) that $R (\vec{k} , \omega)$ cannot be regarded as a structure factor, but as a normalized loss function. In what follows we will discuss the normalized loss function. Notice, that in the classical case the loss function and the dynamic structure factor coincide.

%\begin {figure} [h] 
%\vspace*{-1cm}
%\unitlength1mm
%  \begin{minipage}[t]{7.4cm}
%   \begin{picture}(80,70)
%\put(-3,-9){\psfig{figure=fig2.eps,width=8.0cm,height=7.0cm,angle=0}}
%   \end{picture}\par  
%\caption{\label{G1th1} $\theta=1$}
%\caption{\label{G1th1} Comparison of the MD and RPA loss function $R(q,\omega)$ versus frequency $\omega/\omega_p$ for different wavevectors q at $\Gamma=1$ and $\theta=1$, {\cite{OrScEb97}}.}
%\end{minipage}
%\hfill
%\begin {minipage} [t]{7.4cm} 
%\begin{picture}(80,70) 
%\put(-5,-9){    \psfig{figure=fig3.eps,width=8.0cm,height=7.0cm,angle=0}}
%   \end{picture}\par  
%\caption{\label{G1th50} $\theta=50$}
%\caption{\label{G1th50} same as Fig.2; for $\Gamma=1$ and $\theta=50$, {\cite{OrScEb97}} }
%\end{minipage}\par
%\vspace*{0.5cm}
%{\parbox[t]{154mm}{Figures \ref{G1th1}, \ref{G1th50}: Comparison of the MD and RPA loss function $R(q,\omega)$ versus frequency $\omega/\omega_p$ for different wavevectors q at $\Gamma=1$ and for various $\theta$ {\cite{OrScEb97}}}}
%\end{figure}

\begin {figure} [h] 
\unitlength1mm
  \begin{picture}(120,90)
{\psfig{figure=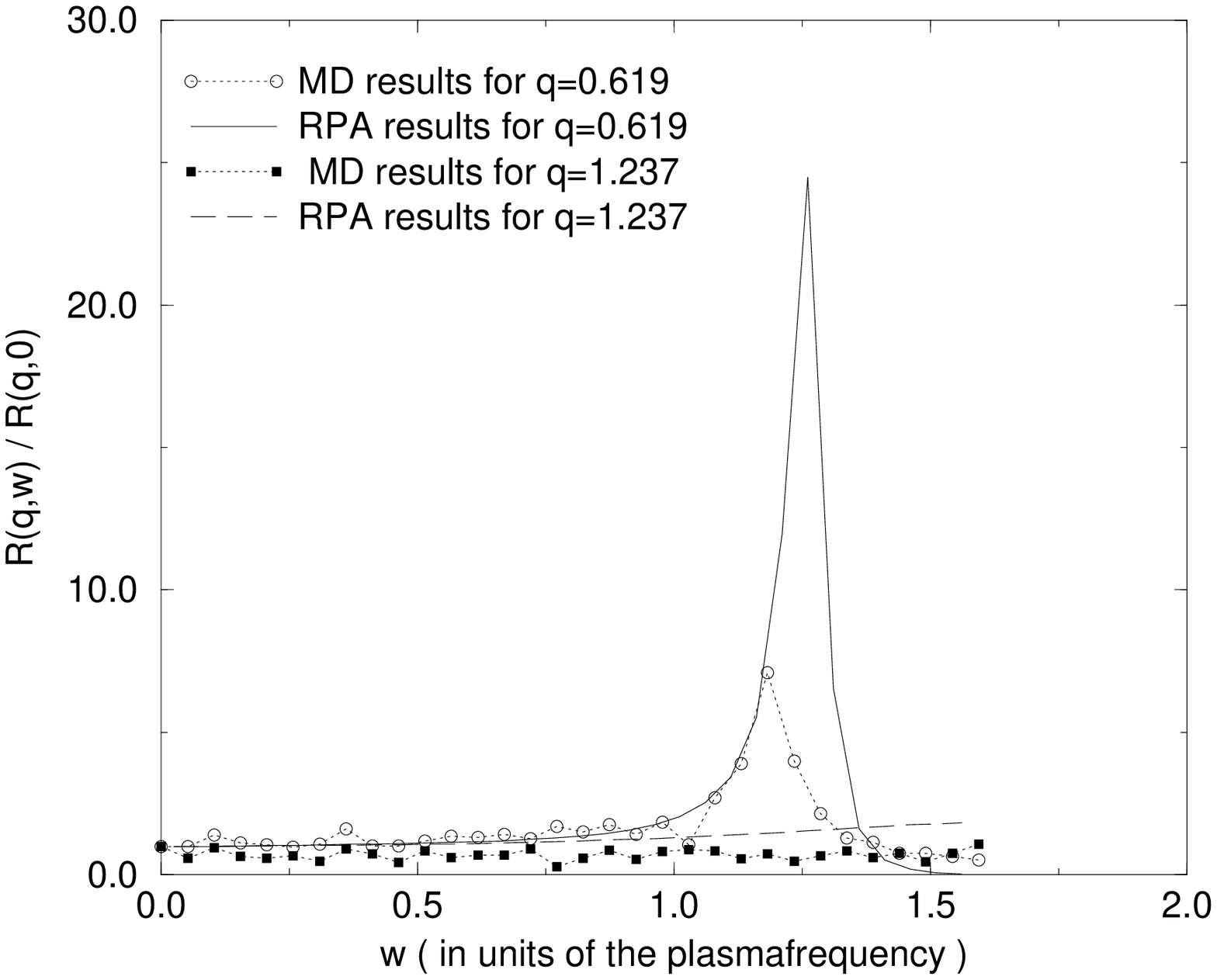,width=12.0cm,height=9.0cm,angle=0}}
 \end{picture}\par 
\caption{\label{G1th1} {Comparison of the MD and RPA loss function $R(q,\omega)$ versus frequency $\omega/\omega_p$ for different wavevectors q at $\Gamma=1$ and $\theta=1$,
 {\cite{OrScEb97}}
.}}
\end{figure}

\begin {figure} [h] 
\unitlength1mm
  \begin{picture}(120,100)
\psfig{figure=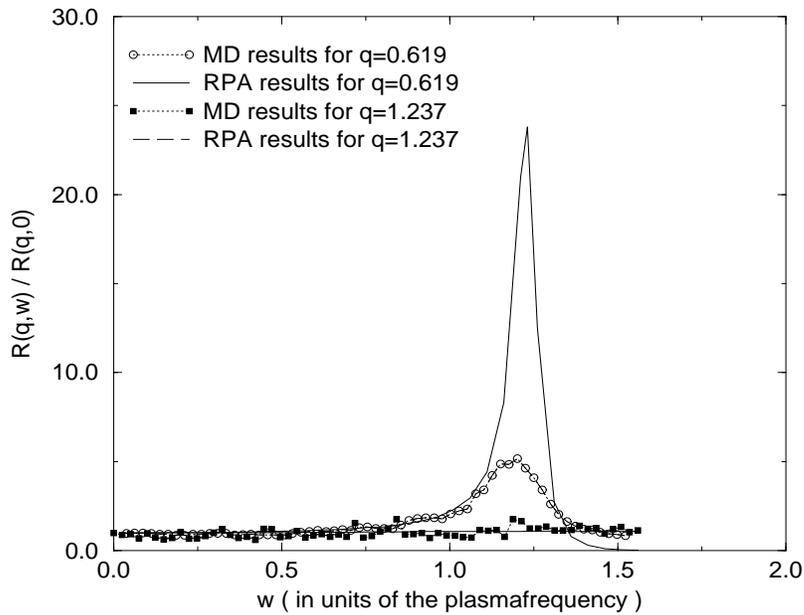,width=12.0cm,height=9.5cm,angle=0}
 \end{picture}\par 
\caption{\label{G1th50} {same as Fig. \ref{G1th1}; for $\Gamma=1$ and $\theta=50$, 
%\cite{OrScEb97}
. }}
\end{figure}

\begin {figure} [h] 
\unitlength1mm
  \begin{picture}(120,100)
\psfig{figure=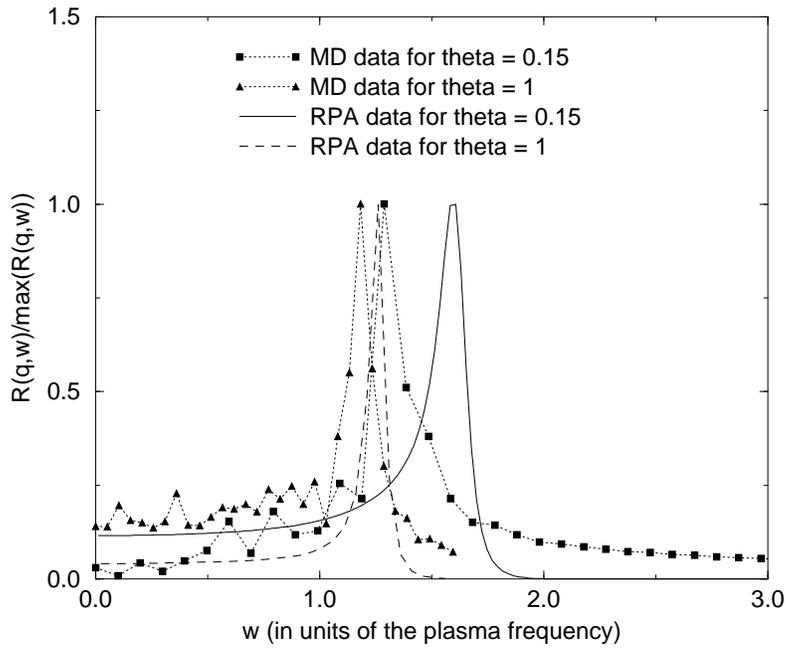,width=12.0cm,height=10.0cm,angle=0}
 \end{picture}\par 
\caption{\label{thvergl}{The MD loss function $R(q,\omega)$ versus frequency $\omega/\omega_p$  for wavevector $q=0.619$ at fixed $\Gamma=1$ and different $\theta$.}}
\end{figure}

\begin {figure} [h]
\unitlength1mm 
  \begin{picture}(120,100)
\psfig{figure=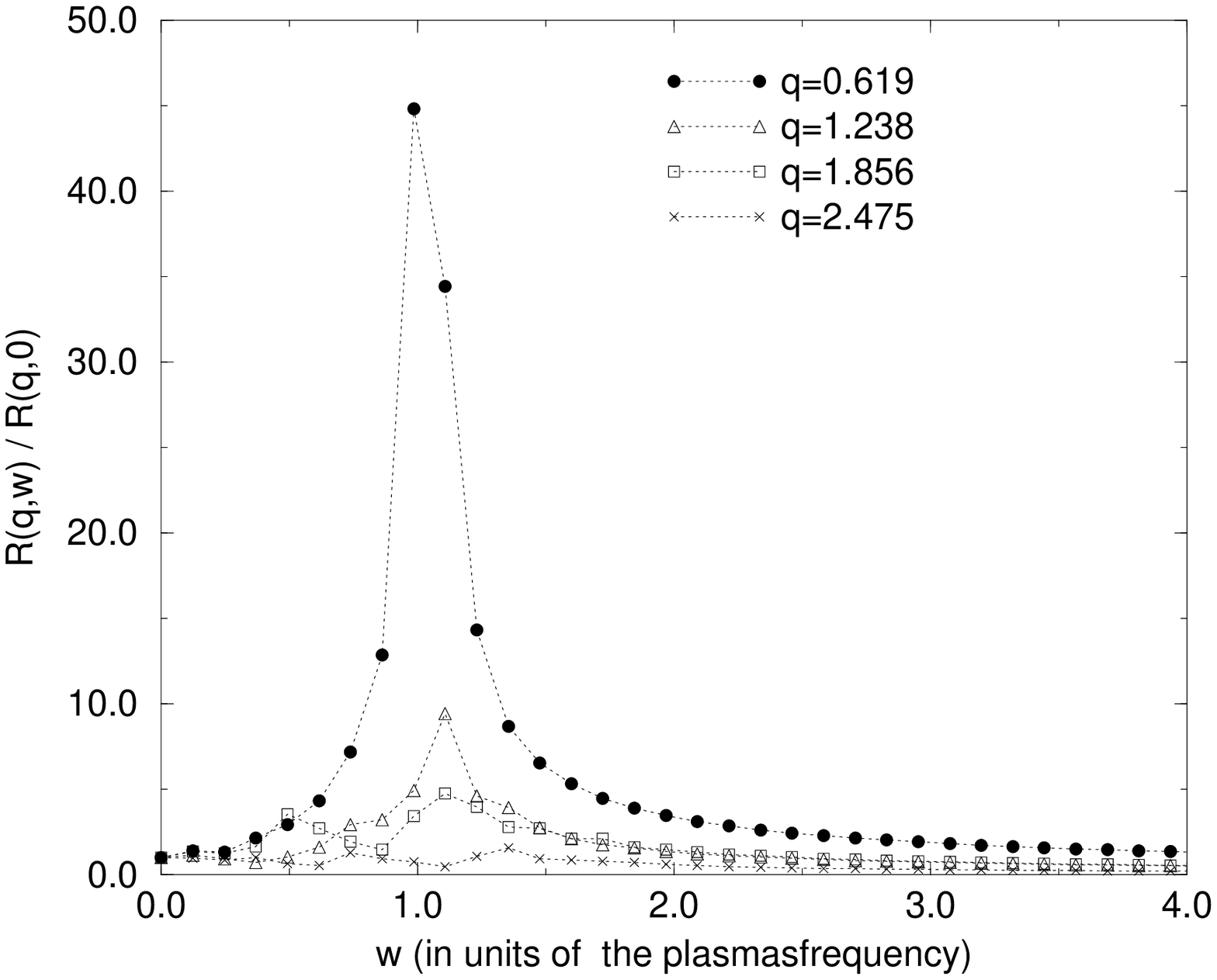,width=12.0cm,height=10.0cm,angle=0}
 \end{picture}\par 
\caption{\label{G10th1} {The MD loss function $R(q,\omega)$ versus frequency $\omega/\omega_p$  for different wavevectors q at $\Gamma=10$ and $\theta=1$.}}
\end{figure}

\begin {figure} [h]
\unitlength1mm 
  \begin{picture}(120,100)
\psfig{figure=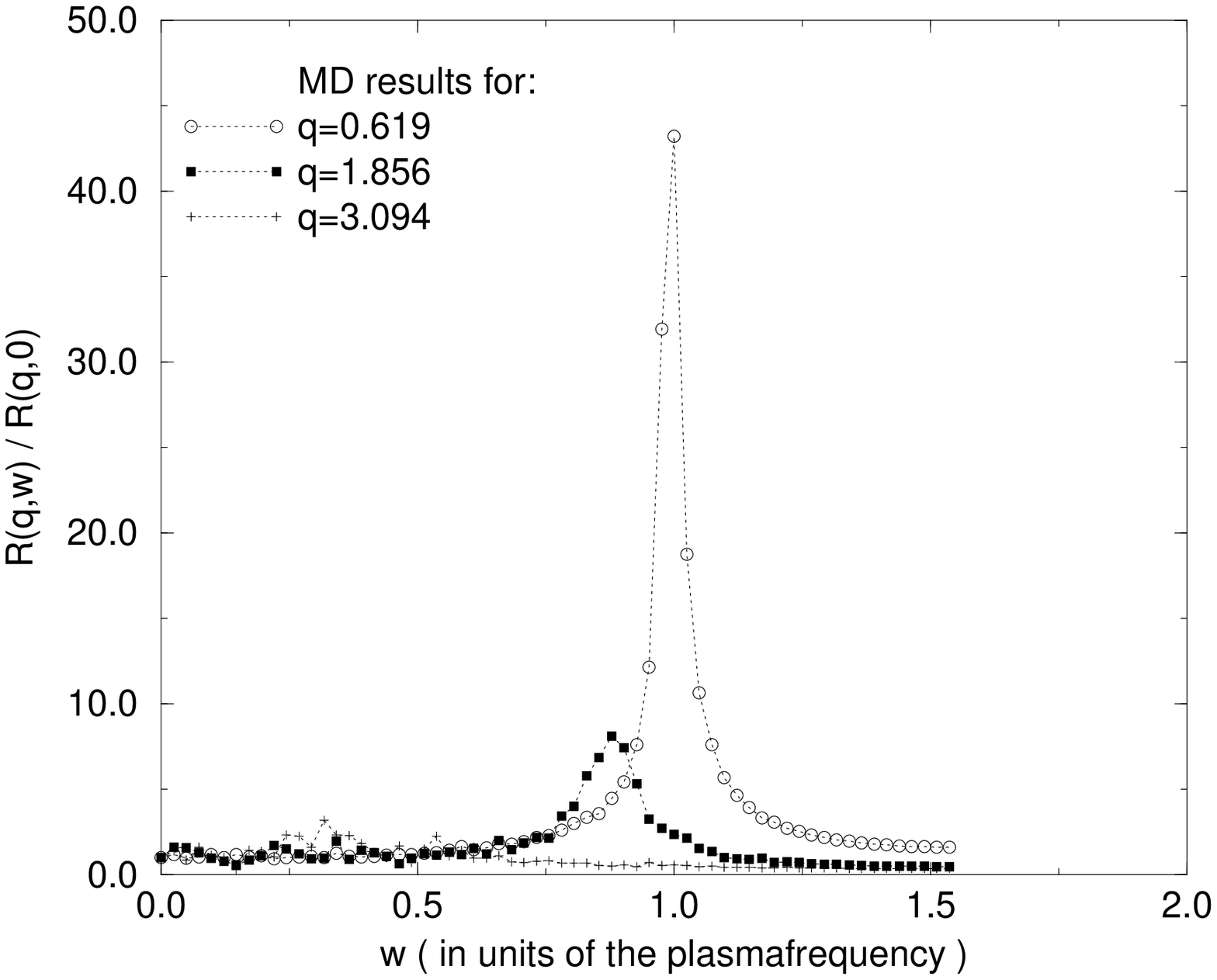,width=12.0cm,height=10.0cm,angle=0}
 \end{picture}\par 
\caption{\label{G100th50} {The MD loss function $R(q,\omega)$ versus frequency $\omega/\omega_p$  for different wavevectors q at $\Gamma=100$ and $\theta=50$}}
\end{figure}

\begin {figure} [h] 
\unitlength1mm
  \begin{picture}(120,100)
\psfig{figure=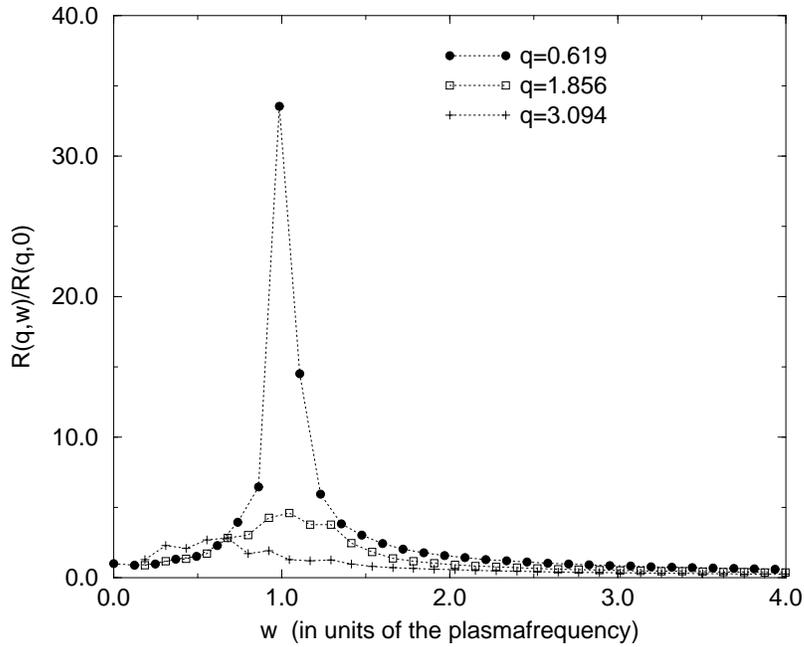,width=12.0cm,height=10.0cm,angle=0}
 \end{picture}\par 
\caption{\label{G100th1} { same as Fig. \ref{G100th50}, at $\Gamma=100$ and $\theta=1$.}}
\end{figure}

We have plotted the loss function $ R(q,\omega) $ ($ q=ka $) for various q values, for the case of moderate coupling ($ \Gamma=1 $), strong coupling ($ \Gamma=10$ ) and for the case of very strong coupling ($ \Gamma=100$ ) and for different parameters of degeneracy $ \theta=1 $ (moderate degenerate) and $ \theta = 50 $ (classical plasma) (Figs. \ref{G1th1}-\ref{G100th1}). For the regime of moderate coupling we have compared the results of the simulations with the corresponding data from the Random Phase approximation (RPA), {\cite{OrScEb97}}.  We see from Figs.  \ref{G1th1}, \ref{G1th50} that in the case of a moderate coupled plasma ($ \Gamma = 1 $) the shape of the loss function calculated from the simulations is damped stronger and slightly shifted to the left in comparison with the RPA loss function. For moderate coupling and in both cases (weakly degenerate and classical plasma) the plasmon peak can be observed only for the smallest q value (q=0.619). From Figs  \ref{G1th1} and \ref{G1th50} it can be also seen that the change of the degeneracy parameter $ \theta $ in the range from 50 to 1 has only a small influence on the results.

%\begin {figure} [h]
%\vspace*{-1cm} 
%  \unitlength1mm
% \begin{minipage}[t]{7.4cm}
%   \begin{picture}(80,70)
%\put(-3,-9){    \psfig{figure=figv.eps,width=8.0cm,height=7.0cm,angle=0}}
%   \end{picture}\par 
%\caption{\label{thvergl} The MD loss function $R(q,\omega)$ versus frequency $\omega/\omega_p$  for wavevector $q=0.619$ at fixed $\Gamma=1$ and different $\theta$}
%\end{minipage}
%\hfill
% \begin{minipage}[t]{7.4cm}
%   \begin{picture}(80,70)
%\put(-5,-9){    \psfig{figure=fig10_1.eps,width=8.0cm,height=7.0cm,angle=0}}
%   \end{picture}\par  
%\caption{\label{G10th1} The MD loss function $R(q,\omega)$ versus frequency $\omega/\omega_p$  for different wavevectors q at $\Gamma=10$ and $\theta=1$}
%\end{minipage} 
%\end{figure} 

However, at higher degrees of degeneracy ($\theta=0.15$) the plasmon peak obtained from the MD datas is shifted towards higher frequencies. But the peak position in this case differs quite significantly from those predicted by the RPA which is shifted much more due to the increase of the average velocity (the Fermi velocity) which leads to a strong positive dispersion (Fig. \ref{thvergl}). The simulations underestimate this shift.

In the strong coupling regime ($ \Gamma = 10 $) and for a moderate degenerate plasma ($ \theta=1 $) we observed at the smallest q value  a sharp plasmon peak centered  near $ \omega_P $ (Fig. \ref{G10th1}).  With the increasing wavevector the plasmon peak widens, a plasmon peak can be observed up to $q=1.856$. At still higher q values the peak vanishes. Notice also that within the resolution limited by the simulations almost no dispersion could be observed at this thermodynamic conditions. We interpretate this as the limiting region from positive (at smaller $\Gamma$) to negative dispersion (at higher $\Gamma$).

%\begin {figure} [h]
%\vspace*{-1cm} 
%  \unitlength1mm
% \begin{minipage}[t]{7.4cm}
%   \begin{picture}(80,70)
%\put(-3,-9){    \psfig{figure=fig4.eps,width=8.0cm,height=7.0cm,angle=0}}
%   \end{picture}\par
%\caption{\label{G100th50} $\theta=50$}  
%\caption{\label{G100th50} The MD loss function $R(q,\omega)$ versus frequency $\omega/\omega_p$  for different wavevectors q at $\Gamma=100$ and $\theta=50$}
%\end{minipage}
%\hfill
% \begin{minipage}[t]{7.4cm}
%   \begin{picture}(80,70)
%\put(-5,-9){    \psfig{figure=fig100_1.eps,width=8.0cm,height=7.0cm,angle=0}}
%   \end{picture}\par 
%\caption{\label{G100th1}$\theta=1$}   
%\caption{\label{G100th1} same as Fig. \ref{G100th50}, at $\Gamma=10$ and $\theta=1$}
%\end{minipage} \par
%\vspace*{0.5cm}
%Figures \ref{G100th50}, \ref{G100th1}: The MD loss function $R(q,\omega)$ versus frequency $\omega/\omega_p$  for different wavevectors q at $\Gamma=100$ for different $\theta$.
%\end{figure} 

In the case of very strong coupling $ \Gamma = 100 $ an extremely sharp plasmon peak centered  near $ \omega_P $ can be observed at the smallest q value   (Figs. \ref{G100th50}, \ref{G100th1}).  Here the plasmon peak can be observed up to $q=3.094$. For both the case of a classical plasma ($\theta=50$) and for the case of moderate degenerate plasma ($\theta=1$) a negative dispersion is seen (with increasing wavenumber the peak position is shifted more and more to the left).  This behavior contradicts to the RPA where no plasmon peak is predicted in this regime due to the strong Landau 
damping . However, the RPA cannot be applied to the strong coupling regime.  On the contrary, the results of our simulations for the case of a weakly degenerate plasma ($\theta=50$) are in a 
good agreement with the results of corresponding MD simulations of Hansen et al. for the classical one component plasma {\cite{hansen}}. From Figs. \ref{G100th50} and \ref{G100th1} one also sees that the negative dispersion is more pronounced in the case of a classical plasma.

%\begin {figure} [h] 
%\vspace*{-1cm}
% \unitlength1mm
%  \begin{minipage}[t]{7.4cm}
%   \begin{picture}(80,70)
%\put(-3,-9){    \psfig{figure=fig6.eps,width=8.0cm,height=7.0cm,angle=0}}
%   \end{picture}
%\caption{\label{Nev1} $q=1.856$}   
%\caption{\label{Nev1} comparison of the MD loss function $R(q,\omega)$ versus frequency $\omega/\omega_p$ with the corresponding loss function from the sum rules approach at $\Gamma=100$ and $\theta=50$ for wavevector $q=1.856$}
%\end{minipage}
%\hfill
%\begin{minipage}[t]{7.4cm}
%   \begin{picture}(80,70)
%\put(-5,-9){    \psfig{figure=fig7.eps,width=8.0cm,height=7.0cm,angle=0}}
%   \end{picture}
%\caption{\label{Nev2} $q=3.094$}   
%\caption{\label{Nev2} same as Fig.6; at $\Gamma=100$ and $\theta=50$ for wavevector $q=3.094$}
% \end{minipage}\par
%\vspace*{0.5cm}
%{Figures \ref{Nev1}, \ref{Nev2}: Comparison of the loss function $R(q,\omega)$ from MD simulations and sum rules approach vs.frequency $\omega/\omega_p$  at $\Gamma=100$ and $\theta=50$ for various wavevector \cite{OrScEb97}.}
%\end{figure}

%\begin {figure} [h] 
%\unitlength1mm
%  \begin{minipage}[t]{15.4cm}
%   \begin{picture}(154,85)
%    \put(25,0){\psfig{figure=fig8.eps,width=9.5cm,height=8.0cm,angle=0}}
%   \end{picture}\par  
%\caption{\label{Nev3} same as Fig.6; at $\Gamma=1$ and $\theta=50$ for wavevector $q=0.619$}
%\end{minipage}
%\end{figure}

\begin {figure} [h]
\unitlength1mm 
  \begin{picture}(120,100)
\psfig{figure=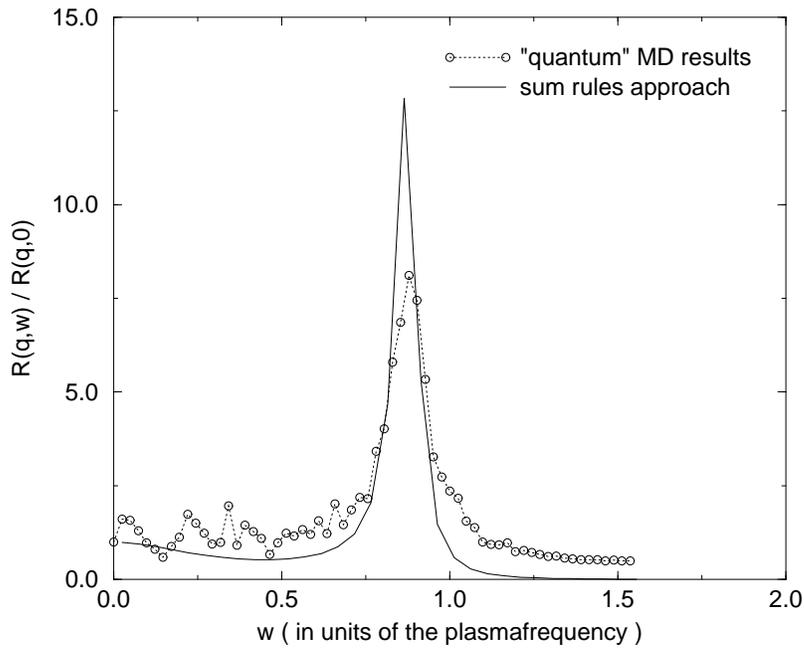,width=12.0cm,height=10.0cm,angle=0}
 \end{picture}\par 
\caption{\label{Nev1} {comparison of the MD loss function $R(q,\omega)$ versus frequency $\omega/\omega_p$ with the corresponding loss function from the sum rules approach at $\Gamma=100$ and $\theta=50$ for wavevector $q=1.856$, 
% {\cite{OrScEb97}}
.}}
\end{figure}

\begin {figure} [h] 
\unitlength1mm
  \begin{picture}(120,100)
\psfig{figure=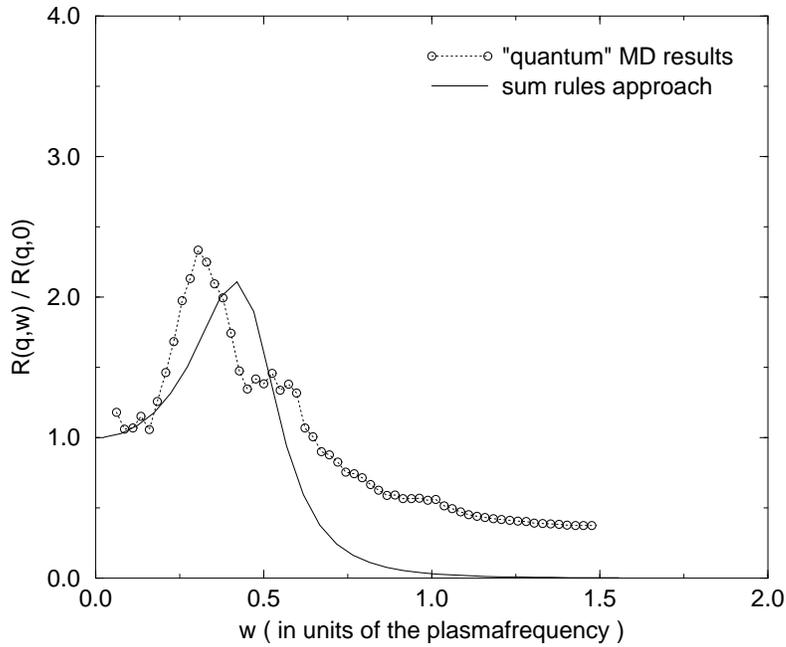,width=12.0cm,height=10.0cm,angle=0}
 \end{picture}\par 
\caption{\label{Nev2} { same as Fig.\ref{Nev1}; at $\Gamma=100$ and $\theta=50$ for wavevector $q=3.094$, 
% {\cite{OrScEb97}}
.}}
\end{figure}

Further we have compared the results of our simulations with the expression of the dynamic structure factor obtained by the application of the classical theory of moments which is appropriate also for the case of strong coupling {\cite{Adamyan}}. A more detailed discussion can be found elsewhere {\cite{OrScEb97}}. Here we show the results for two different q vectors at $\Gamma=100$ (Figs.\ref{Nev1}, \ref{Nev2} \cite{OrScEb97}). The agreement of the loss functions calculated by the application of the theory of moments (sum rules approach) with that from the MD calculations is rather good. The theoretical curves reproduce the varying shape of the dynamic structure factor and describe the plasmon peak position in a satisfactory manner. However, the agreement in the height of the peaks is less satisfactory .
% One of the reasons for this disagreement between the results of simulations with theoretical predictions (both based on RPA and on the sum rules approach) might be the normalization to $S(q,0)$ which is a value rather bad measured in the simulations due to the poor statistics at long times.

Thus we can conclude that due to the quasi-classical character our quantum molecular dynamic simulations describe  the collective excitations of the electron gas only approximately. Our model yields quite reasonable results at weak and moderate degeneracy, whereas for the case of high degeneracy it seems to break down.

\section{Discussion} 

We developed here a simple quasi-classical model
of quantum plasmas based on a dynamics with an effective momentum-dependent
Hamiltonian.
The quantum-mechanical effects corresponding to the Pauli and the
Heisenberg principle were modeled by constraints in the Hamiltonian.
By using the concept of minimum uncertainty wave packets,
momentum-dependent effective potentials were derived.
The effective potentials were used to simulate one-component plasmas and
mass -symmetrical two - component plasmas by means of MC and MD methods. 
The result of the simulations is in good 
agreement with analytical calculations of the thermodynamic properties in the
region of small degeneracy and moderate coupling. 

MD studies provide also informations on the dynamical properties of the plasma,
in particular on collective excitations. As a basis to study these effects we
calculated the structure factor of the plasma. The agreement between the
simulations and the analytical theory based on RPA, sum rules and the theory 
of moments is reasonable. In most cases the shape and the location of the plasmon peak is
reproduced in a reasonable way, in some cases larger deviations are observed.
We cannot expect however, that the present model, which was obtained
by fitting the potential to equilibrium properties will describe
all non-equilibrium properties in a quantitative way. Further improvements
of the model might be unavoidable. We may hope however that at least near
to equilibrium some realistic features are still reflected by the model.

Acknowledgment: The authors thank V.M. Adamyan, Yu.M. Klimontovich,
B. Militzer, V. Podlipchuk and I.M. Tkachenko for helpful discussions.

\vspace*{-1cm}
%\newpage

\newpage

%\newpage

\newpage

%\newpage

\newpage

%\newpage

\end{document}